\documentclass[a4,11pt]{article}
\usepackage{amsmath}
\usepackage{graphicx}
\usepackage{amssymb} 
\newcommand{\Mat}[1]{{{\boldsymbol{#1}}}}
\newcommand{\abs}[1]{\left\vert#1\right\vert}
\def\be{\begin{equation}}
\def\ee{\end{equation}}
\def\dd{\mathrm{d}}
\title{Scalar gravity with preferred frame: asymptotic post-Newtonian scheme and the weak equivalence principle}

\bigskip
\author{
Mayeul Arminjon \\
\small\it Laboratoire ``Sols, Solides, Structures'' (Unit\'e Mixte de Recherche of the CNRS), \\[-1.mm]
\small\it BP 53, F-38041 Grenoble cedex 9, France
}
\date{ }

\begin{document}

\maketitle

\begin{abstract}
A scalar theory of gravity with a preferred reference frame is presented. It is insisted on the dynamics, which involves a (non-trivial) extension of Newton's second law, and on the new version (``{\bf v2}") with isotropic space metric. We display the energy conservation equation obtained with {\bf v2}. Then the principles of the asymptotic post-Newtonian approximation are discussed in some detail. The results of its application to the motion of a small extended body in a weakly-gravitating system are given and discussed: the weak equivalence principle was violated in {\bf v1}, due to its anisotropic space metric (as the standard Schwarzschild metric), but is valid with {\bf v2}.

\end{abstract}

\bigskip

\section{Introduction}

A preferred-frame theory of gravitation, based on just a scalar field, has been previously proposed and studied by this author. It is derived from a semi-heuristic interpretation of gravity as due to a pressure gradient in a universal fluid or ``ether", but it consists of a definite and consistent set of equations; see Ref. \cite{O2} for a detailed motivation and a summary of the construction and the test of that theory. Although this theory has passed a number of tests, it has recently been discarded by a surprising violation of the weak equivalence principle (WEP), which has been found to occur for {\it real bodies} of a small but finite size \cite{A33}, whereas the WEP is a built-in feature of the theory for {\it test particles}. That violation can be seen, due to the use of a rigorous approximation method for weak gravitational fields, which the simplicity of this theory makes it possible to implement. \\

It turns out that the theory can be saved by modifying the space metric and the equation for the scalar field, without changing the semi-heuristic foundation nor the dynamical equations---which are the two characteristic features of the theory. The main aim of this contribution is to present the new version (``{\bf v2}") and the asymptotic scheme of approximation in a self-contained way. It shall be insisted on the foundations of both the theory and the approximation method, but many calculations shall be skipped. The reasons why the violation of the WEP occurs in {\bf v1} and not in {\bf v2} shall be exposed. Some comparisons with general relativity (GR) shall also be inserted.

\section{Presentation of the scalar ether-theory (v1 \& v2)} \label{Principles}

\subsection{Preferred reference frame and space-time metric} \label{PRF}

Space-time is assumed to be the product V $=\mathsf{R} \times \mathrm{M}$, where $\mathrm{M}$ is the preferred reference body (as Newton's absolute space), endowed with an Euclidean metric $\Mat{g}^0$. The equations of the theory are primarily written in the preferred reference frame E in which the body M is at rest. [A general reference frame is for us essentially a reference body N, plus a notion of time. As to the frame E, associated with the body M, it is endowed with the ``absolute time" $T \equiv x^0/c$, where $x^0$ is the canonical projection of $X \in$ V into $\mathsf{R}$, and $c$ is a constant---the limit velocity of mass points (see point \ref{limit-c}).] Thus, we shall usually formulate equations that are space-covariant only. The Euclidean metrics on the component spaces $\mathsf{R}$ and M allow to define a flat Lorentzian metric $\Mat{\gamma}^0$ : for a 4-vector $\mathsf{U}=(U^0,{\bf u})$ (with $U^0$ a real number and ${\bf u}$ a spatial vector, i.e. an element of the tangent space TM$_{\bf x}$ to M at some ${\bf x} \in M$),
\begin{equation} \label {flatspacetimemetric} 
\Mat{\gamma}^0(\mathsf{U},\mathsf{U})=(U^0)^2-\Mat{g}^0({\bf u},{\bf u}).
\end{equation}
This metric would measure the proper time, if there were no gravity. The gravity field is a scalar field $\beta(T, {\bf x})$, that both has metrical effects {\it and} produces a gravity acceleration. The physical space-time metric $\Mat{\gamma}$ is related to the flat metric $\Mat{\gamma}^0$ through the scalar field $\beta $:
\begin{equation} \label {spacetimemetric} 
\Mat{\gamma}(\mathsf{U},\mathsf{U})=\beta^2(U^0)^2-\Mat{g}({\bf u},{\bf u}),
\end{equation}
with $\Mat{g}$ the physical space metric on the preferred body M, which is itself related to $\Mat{g}^0$ through the scalar field $\beta$ as explained in Subsect. \ref{SpaceMetric}. Equation (\ref{spacetimemetric}) implies that the scalar $\beta$ can be defined as
\be \label{beta_operatoire}
\beta \equiv (\gamma_{00})^{1/2},
\ee
in any coordinates $(y^\mu)$ {\it adapted} to the frame E [i.e., such that each point ${\bf x} \in \mathrm{M}$ has constant space coordinates $(y^i)\ (i=1,2,3)$], and such that $y^0 = x^0 \equiv cT$. From (\ref{spacetimemetric}), it also follows that $\gamma _{0i}=0 \ (i=1,2,3)$ in such coordinates. 

\subsection{Dynamics} \label{Dynamics} 
\subsubsection{The gravity acceleration in the scalar ether-theory}
An important equation of this theory is that for the gravity acceleration:
\be \label{g_beta}
\mathbf{g} \equiv -c^2\frac{\mathrm{grad}_{\Mat{g}}\beta}{ \beta}, 
\ee
where the gradient vector with respect to the physical, Riemannian space metric $\Mat{g}$ is defined by the space-contravariant components $(\mathrm{grad}_{\Mat{g}}\beta)^i=g^{ij}\beta_{,j}$, with $(g^{ij})$ the inverse matrix of matrix $(g_{ij})$. This equation can be obtained in two different ways. Let us outline these two ways. The {\it first way} is semi-heuristic: one may interpret gravity as {\it Archimedes' thrust} due to the gradient of the macroscopic (smoothed-out) pressure $p_e$ in an imagined fluid called ``micro-ether", of which the elementary particles would be local organizations (e.g. vortices). This gives \cite{O3}
\be \label{g_rho_e}
\mathbf{g}=-\frac{\mathrm{grad}\, p_e}{\rho_e},
\ee
where $\rho _e=\rho _e(p_e)$ is the macroscopic ``ether density", which thus {\it decreases} towards the gravitational attraction. Then, in that framework, the Lorentz-Poincar\'e interpretation of special relativity (SR) \cite{Prokhovnik67,Prokhovnik93,Brandes01} leads naturally, as summarized in Subsect. \ref{SpaceMetric}, to assume \cite{O3} that there are metrical effects of a gravitational field, which depend on the ratio 
\be \label{beta-rho_e}
\beta = \rho _e/\rho _e^\infty \leq 1,
\ee
where 
\be \label{rho_e^infty}
\rho _e^\infty(T)\equiv \mathrm{Sup}_{\mathbf{x}\in\mathrm{M}}\ \rho_e(\mathbf{x},T) 
\ee
is the ether density in remote regions that are free from gravitational field. [Precisely, this assumption leads to $\sqrt{\gamma_{00}}=\rho _e/\rho _e^\infty$, thus Eq.~(\ref{beta-rho_e}) is consistent with Eq.~(\ref{beta_operatoire}).] Furthermore, the limiting velocity $c$ of SR is then interpreted as the ``sound" velocity in the ether \cite{O3}, thus giving 
\be \label{p_e-rho_e}
p_e=c^2 \rho _e.
\ee
Equations (\ref{beta-rho_e}) and (\ref{p_e-rho_e}), combined with (\ref{g_rho_e}), lead indeed to Eq. (\ref{g_beta}). The {\it second way} to obtain Eq. (\ref{g_beta}) is as follows: as recalled below, one may uniquely define Newton's second law in any relativistic theory of gravity with curved space-time, the gravitational force being $m(v){\bf g}$ with $m(v) \equiv m(0) \gamma _v$ the relativistic inertial mass and ${\bf g}$ a theory-dependent gravity acceleration. ($v \equiv \Mat{g}({\bf v},{\bf v})^{1/2}$ is the modulus of the velocity ${\bf v} $, and $\gamma _v$ is the Lorentz factor.) Now, $\beta$ being defined by Eq. (\ref{beta_operatoire}), postulating Eq. (\ref{g_beta}) for vector ${\bf g}$ is {\it equivalent}, under natural requirements, to {\it asking geodesic motion in the case of a static gravitational field} \cite{A16}.

\subsubsection{Newton's second law in a curved space-time}
To use a gravity acceleration, we need to define dynamics of a test particle by extending the special-relativistic form of Newton's second law to the case with gravitational force in a curved space-time---instead of postulating that free test particles follow space-time geodesics, as is done in GR. This can be done for any theory with Lorentzian space-time metric. The gravitational force has just been introduced, and the momentum must of course be 
\be \label{momentum}
\mathbf{P} \equiv m(v)\,\mathbf{v}, 
\ee
where the velocity ${\bf v}$ has to be defined in terms of the ``local time" $t_\mathbf{x}$, measured by a clock at the spatial position $\mathbf{x} \equiv (y^i)$ that is fixed in the frame F considered, and that momentarily coincides with the position of the test particle---thus 
\be
{\bf v} \equiv \dd {\bf x}/\dd t_\mathbf{x}. 
\ee
\{In any theory, the frame F is fixed by the choice of the space-time coordinate system $(y^\mu)$, though many coordinate systems correspond to the same frame. The corresponding body N is described by the space coordinates $(y^i)$, which can be subjected to purely spatial changes $y'^i=\psi^i(y^j)$ \cite{A16,L&L,Cattaneo}.\} For the scalar ether-theory, we consider only the preferred reference frame, thus F = E, and we have from (\ref{spacetimemetric}):
\begin{equation} \label {localtime}
		 			\dd t_\mathbf{x}/\dd T = \beta(T,\mathbf{x}).
\end{equation}
However, a synchronized local time may also be defined for any trajectory $\xi \mapsto X(\xi)\equiv (y^\mu(\xi))$ in the arbitrary reference frame F, in any theory with curved space-time metric \{see Landau \& Lifshitz \cite{L&L}, Eqs. (84,14) and (88,10), and Cattaneo \cite{Cattaneo}\}: 
\begin{equation} \label {localtimegeneral}
		 			\frac{\dd t_\mathbf{x}}{\dd \xi } = \frac{\sqrt{\gamma _{00}}}{c} \left( \frac{\dd y^0}{\dd \xi } + \frac{\gamma _{0i}}{\gamma _{00}} \frac{\dd y^i}{\dd \xi } \right).
\end{equation}
To complete the definition of Newton's second law, we have to define the time-derivative of a spatial vector (the momentum ${\bf P}$ of the test particle in the frame F). This is not trivial in the general case of a variable gravitational field, which means a time-dependence of the spatial metric $\Mat{g}$ in the frame F. We have found \cite{A16} that a unique definition may be given for the time-derivative $D\mathbf{w}/D\xi$ of a space vector $\mathbf{w}$ depending on a parameter $\xi$ along a trajectory in a space (manifold) N endowed with a metric $\Mat{g} = \Mat{g}_\xi$ that varies with the parameter $\xi$, under compelling requirements that include the validity of Leibniz' differentiation rule for a scalar product. In coordinates, this unique definition is as follows:
\begin{equation} \label{timederivative}
\left(\frac{D\mathbf{w}}{D\xi}\right)^i \equiv \frac{\dd w^i}{\dd \xi}\,+ \Gamma^i_{jk} w^j\,\frac{\dd y^k}{\dd \xi} + \frac{1}{2} {\mathsf t^i_j} w^j,\quad \Mat{t} \equiv \Mat{g}_\xi ^{-1}\mathbf{.}\frac{\partial \Mat{g}_\xi } {\partial \xi}, 
\end{equation}
with $\Gamma^i_{jk}$ the Christoffel symbols of metric $\Mat{g}$. Our extension of Newton's second law is hence:
\begin{equation} \label{Newtonlaw}
m(v)\,\mathbf{g} = \frac{D\mathbf{P}}{Dt_\mathbf{x}},
\ee
thus the parameter $\xi$ in (\ref{timederivative}) is then the synchronized local time, given in general by Eq.~(\ref{localtimegeneral}), and in the investigated theory by Eq.~(\ref{localtime}). It turns out that Einstein's motion following the geodesic lines of $\Mat{\gamma }$ {\it can be put in the form (\ref{Newtonlaw}):} if the coordinate system can be chosen such that $\gamma _{0i}=0 \ (i=1,2,3)$ (which is true for a class of possible reference frames), the gravity acceleration has to be \cite{A16}
\be \label{g-geod}
{\bf g}_\mathrm{geod} = -c^2\frac{\mathrm{grad}_{\Mat{g}}\beta}{\beta}- \frac{1}{2} \Mat{t}{\bf .v},
\ee
where $\beta \equiv \sqrt{\gamma _{00}}$ and $\Mat{g}$ is the space metric in the reference frame fixed by the coordinate system. This is invariant under the coordinate transformations \cite{A16} that leave the reference frame unchanged.

\subsubsection{The limit velocity and the case with photons. The geodesic motion} \label{limit-c}
It follows from Eq.~(\ref{Newtonlaw}) that a ``mass point" (a test particle with positive rest-mass) cannot reach the velocity $c$. Hence, that velocity is indeed an upper limit for mass points, as in special relativity. For the investigated theory, this limit applies a priori in the ether frame E, however, because the equations for the gravity acceleration (\ref{g_beta}) and for the metric (\ref{spacetimemetric}) apply in that frame. For a photon, the same law is used, substituting in Eqs. (\ref{momentum}) and (\ref{Newtonlaw}) the mass content $E/c^2$ of the energy $E=h\nu$ for the inertial mass $m(v)$, the frequency $\nu$ being evaluated with the local time. As applied to photons, our extension of Newton's second law implies that they keep the constant velocity $c$ (again in the frame E) \{see Eq. (29) in Ref. \cite{A34}\}. Now, when going to an arbitrary reference frame, we transform the space-time metric [defined in the frame E by Eq.~(\ref{spacetimemetric})] as a space-time tensor, of course. This actually means that we are using the standard (Poincar\'e-Einstein) synchronization convention of clocks \{see Ref. \cite{L&L}, Sect. 84\}. Thus the time-like, light-like, or space-like character of a world line element is, of course, invariant. In other words, photons will also run at the velocity $c$ in any physically possible reference frame, but, in frames other than the ether frame, this has partly a conventional character: only the {\it back-and-forth} velocity of light is physically meaningful \cite{Prokhovnik67,Brandes01,Selleri96}. As already mentioned, it also follows from Newton's second law (\ref{Newtonlaw}) with the gravity acceleration (\ref{g_beta}) that, {\it in the case of a static gravitational field} (which, in that theory, is characterized by condition $\beta _{,0}=0$), test particles follow geodesic lines of metric $\Mat{\gamma }$. This is true for mass points \cite{A16} [as can be seen immediately by comparing Eqs.~(\ref{g_beta}) and (\ref{g-geod})] and for photons as well \cite{A18}.

\subsubsection{Continuum dynamics and Maxwell equations}
To describe the {\it motion of a continuous medium} (be it a fluid or an {\it electromagnetic field}, for instance), we begin by investigating the case of a {\it dust}, i.e., a continuum made of non-interacting mass particles. 
For a dust, Newton's second law (\ref{Newtonlaw}) with the gravity acceleration (\ref{g_beta}) may be applied pointwise and implies \cite{A20} that the material energy-momentum tensor ${\bf T}$ verifies the following equation:
\begin{equation} \label{continuum}
T_{\mu;\nu}^{\nu} = b_{\mu},				          
\end{equation}
where
\begin{equation} \label{definition_b}
b_0(\mathbf{T}) \equiv \frac{1}{2}\,g_{jk,0}\,T^{jk}, \quad b_i(\mathbf{T}) \equiv -\frac{1}{2}\,g_{ik,0}\,T^{0k}. 
\end{equation}
[Indices are raised and lowered with metric $\Mat{\gamma}$, unless mentioned otherwise. Semicolon means covariant differentiation using the Christoffel connection associated with metric $\Mat{\gamma}$. Note that in GR, in contrast, we have $b_\mu=0$ in (\ref{continuum}).] The universality of gravity and the mass-energy equivalence mean that the same equation (\ref{continuum}) must hold true, with the same definition (\ref{definition_b}), for any continuous medium. In particular, Eq.~(\ref{continuum}) may be applied to the situation with both a charged medium and the corresponding electromagnetic (e.m.) field. It is in that way that the gravitational modification of the Maxwell equations is got in the scalar ether-theory \cite{B13}. This modification is fully consistent \cite{B13} with photon dynamics as defined [with $E/c^2$ in the place of $m(v)$] by Eq.~(\ref{Newtonlaw}) with (\ref{g_beta}). The wave equivalent of free light-like particles are the {\it null} e.m. fields in vacuo. The consistency means that the trajectories of the energy flux of such fields coincide with the trajectories of free light-like particles \cite{B13}.  This entails, in particular, the prediction that, {\it for null e.m. fields in vacuo}, the energy flux has velocity $c$. When the e.m. field is not a null field, it is not immediate how to define the velocity of the energy flux. It is worth mentioning that Nimtz and his team claim to have measured genuinely superluminal signal velocities in ``photonic tunneling" experiments \cite{NimtzHaibel}. 

\subsection{Specific form for the spatial metric} \label{SpaceMetric}

The foregoing dynamics holds {\it independently of any specific form for the spatial metric} $\Mat{g}$ in the frame E, that enters Eq.~(\ref{spacetimemetric}). If one follows the semi-heuristic analysis of gravity as due to the macroscopic part of the pressure force in the ether, outlined at the beginning of Subsect. \ref{Dynamics}, then one expects a gravitational ``rod contraction" in the {\it same ratio} $\beta \equiv \sqrt{\gamma_{00}} $ as the gravitational ``clock retardation" \cite{O3,A9}. In short, if one analyses Special Relativity within that concept of a barotropic ether, one finds that an observer moving at velocity $u$ with respect to E should consider that, due to the Lorentz contraction, the ether density is smaller for her than for a fixed observer, $\rho'_e=\rho_e/\gamma_u$. Thus, the Lorentz contraction of her meters, as well as the Larmor dilation of the period of her clock, depend on precisely the ratio of the ether density in her moving frame to that in the ether frame. A gravitational field, characterized (according to this semi-heuristic analysis) by a decrease of the density $\rho _e$ in the direction of the gravity acceleration, should hence also lead to rod contraction and clock retardation in the same ratio $\beta \equiv \sqrt{\gamma_{00}}=\rho_e/\rho_e^\infty $. I.e., the {\it physical} ($\Mat{g}$) and {\it Euclidean} ($\Mat{g}^0$) space metrics on the preferred body M should be related together by a dilation of physically-measured distances, with respect to Euclidean distances (due to the contraction of objects, including length standards). But that dilation can occur:
\begin{itemize}
\item {\it either} in a locally-isotropic way, which means postulating the following relationship between the two spatial metrics \cite{O3}:
\begin{equation} \label{spacemetric}  
\qquad\qquad\Mat{g} =	\beta^{-2}\Mat{g}^0 \qquad  ({\bf v2}) .
\end{equation}
\item {\it or}, as in SR, in just one direction---which then can only be that of the gravity acceleration ${\bf g}$. This was the assumption set \cite{A9} in the first version (``{\bf v1}") of the scalar ether-theory, which passed a number of tests \cite{A34,O1}, but which seems to be discarded by the violation of the weak equivalence principle discussed in Sect. \ref{PointParticleLimit} below.
\end{itemize}
Therefore, the author is now investigating in detail the new version (``{\bf v2}"), based on Eq.~(\ref{spacemetric}). 

\subsection{The question of energy conservation}
The conservation of energy is at the very top of the hierarchy of the physical laws. Unfortunately, there is no exact local conservation law for energy in GR. This is due to the fact that energy is not a generally-covariant concept. Essentially, the equation for continuum dynamics in GR, $T_{\mu;\nu}^{\nu} = 0$, may be rewritten as the cancellation of the 4-divergence of a tensor-like object with respect to a {\it flat} metric---i.e., it may be rewritten as an exact local conservation law. {\it But} the equation obtained thus is covariant only under linear coordinate changes, and, moreover, the corresponding ``gravitational energy-momentum pseudo-tensor" is not unique (see e.g. Stephani \cite{Stephani}; see Ref. \cite{A15} for this author's opinion and some other references). \\

Since the theory investigated in this paper is unshamedly a preferred-frame theory, one should be able to ask for a local energy conservation in this theory. To investigate this question, we rewrite the time component of the equation for continuum dynamics (\ref{continuum}) in terms of usual derivatives:
\begin{equation} \label{flat_0}
\left(\sqrt{-\gamma}\, T_0^j\right)_{,j} + \left(\sqrt{-\gamma}\, T_0^0\right)_{,0} =  \sqrt{-\gamma}\, \beta \beta_{,0}\,T^{00}, 
\end{equation}
where
\begin{equation} \label{gamma} 
\qquad \gamma \equiv \mathrm{det}(\gamma_{\mu \nu}) = -\beta^2 \mathrm{det}(g_{ij}).\\
\end{equation}
On the l.h.s. of Eq.~(\ref{flat_0}), we have the energy balance of matter, whereas on the r.h.s. we have a source term depending on the matter energy density (in the preferred frame), $T^{00}$, and on the scalar gravitational field, $\beta\equiv \sqrt{\gamma_{00}}$. Note that, in the sought-for equation for the scalar gravitational field, $\beta $ might be replaced by some function of it, for convenience. This function and the equation for the scalar field are not fully constrained by the heuristic principles of the ether theory \cite{O3}. Another constraint will thus be that, by virtue of this scalar field equation, one should be able to transform the r.h.s. of (\ref{flat_0}) to a 4-divergence, in order to get a {\it conservation equation} for a (material plus gravitational) energy. Due to Eq. (\ref{gamma}), this depends on the form assumed for $\Mat{g}$.

\subsection{The equation for the scalar field and the energy conservation} \label{field&energy}
The heuristic interpretation of gravity as Archimedes' thrust in a compressible fluid leads \cite{O3} to postulate the following general form for the equation governing the scalar gravitational field: 
\be \label{possibleField}
\Delta_{\Mat{g}}\, \rho _e + (\mathrm{time\ derivatives}) =  \frac{4 \pi G}{c^2} \sigma \rho_e F(\beta),  \qquad F(\beta) \rightarrow  1  \mathrm{\ as\ }  \beta \rightarrow 1,                                     
\ee
where $\Delta_{\Mat{g}}$ is the Laplace-Beltrami operator, i.e., the Laplacian associated with the Riemannian metric $\Mat{g}$. The ``time derivatives" term should be such that, in the appropriate limit, involving the condition $\beta \rightarrow 1$ uniformly, the operator on the l.h.s. should become equivalent to the usual (d'Alembert) wave operator. [As is seen on Eqs.~(\ref{spacetimemetric}) and~(\ref{spacemetric}), $\beta \rightarrow 1$ means that the space-time metric $\Mat{\gamma}$ tends towards the flat one $\Mat{\gamma}^0$.] On the r.h.s. of (\ref{possibleField}), $\sigma$ denotes the relevant mass-energy density, which has to be defined in terms of the energy-momentum tensor $\mathbf{T}$. Obviously, this nonlinear wave equation is not fixed. In the first version of the theory, a definite equation had been postulated \cite{A9}, involving the particular condition $F(\beta )=1$. Moreover, it had been found \cite{A15} that the assumed equation could be rewritten in terms of the {\it flat} metric $\Mat{\gamma }^0$ and the field $f \equiv \beta^2$, and did imply a local energy conservation.\\

With the space metric (\ref{spacemetric}) assumed in {\bf v2}, one finds that the term $\Delta_{\Mat{g}}\, \rho_e$ in Eq.~(\ref{possibleField}) can be reexpressed in terms of the Euclidean metric, namely as $-\rho_e^\infty \, \beta^3 \, \Delta_{\Mat{g}^0} \psi $, with $\psi \equiv -\mathrm{Log}\,\beta$. And one finds that the r.h.s. of Eq.~(\ref{flat_0}) can be rewritten as $-\psi _{,0}T^{00}$. Moreover, if one assumes the following {\it linear wave equation} for the scalar field [corresponding to $F(\beta)=\beta^2$ in Eq. (\ref{possibleField})]:
\begin{equation}\label{field} 
\square \psi \equiv \psi _{,0,0}-\Delta \psi = \frac{4\pi G}{c^2} \sigma 
\end{equation}
with 
\be
\sigma \equiv T^{00},
\ee 
then one obtains a local conservation equation for the energy:
\begin{equation} \label{energyconservation}
\partial_0(\varepsilon_\mathrm{m} + \varepsilon_\mathrm{g})+\mathrm{div}({\bf \Phi}_\mathrm{m}+ {\bf \Phi} _\mathrm{g}) = 0. 
\end{equation}
In Eq.~(\ref{energyconservation}), the material and gravitational energy densities are given (in mass units) by:
\be \label{energydensity}
\varepsilon_\mathrm{m} \equiv T^{00}, \qquad \varepsilon_\mathrm{g} \equiv \frac{c^2}{8\pi G} \left[\psi_{,0}^2+\left(\mathrm{grad}\psi \right)^2\right],
\ee
and the corresponding fluxes are:
\be \label{energyflux}
 {\bf \Phi}_\mathrm{m}\equiv  (T^{0j}), \qquad {\bf \Phi}_\mathrm{g} \equiv -\frac{c^2}{4\pi G}  \left(\psi_{,0}\,\mathrm{grad}\psi \right).
\ee
(Henceforth, the operators grad, div and $\Delta$ shall be relative to the Euclidean metric $\Mat{g}^0$.) We do assume Eq.~(\ref{field}) as the equation for the scalar gravitational field in {\bf v2}. This equation and the local energy conservation (\ref{energyconservation}) are invariant under arbitrary changes of {\it spatial} coordinates, but need that the time coordinate is the preferred one, $x^0=cT$. Thus, these are preferred-frame equations. The energy conservation (\ref{energyconservation}) {\it substitutes} for the mass conservation: independently of the assumed form for the space metric, and thus in {\bf v2} as well as in {\bf v1}, one may indeed show that the dynamical equations (\ref{continuum}) imply a reversible creation/destruction of matter in a variable gravitational field \cite{A20}. However, the theory says that mass is extremely close to be conserved in usual situations \cite{A20}.

\section{Asymptotic post-Newtonian approximation (PNA)}
\subsection{Principle of an asymptotic scheme for the field equations of gravitation}

Very accurate experimental comparisons are aimed at in the theory of gravitation, hence one should use a very clean approximation scheme. The purpose of the PNA is to obtain {\it asymptotic expansions} of the fields as functions of a small parameter $\lambda $ (the field-strength). A clean way to obtain this, in accordance with the general principles of asymptotic analysis, is to deduce a family of systems (fields), $(\mathrm{S}^\lambda)$, from the data of the system of interest, S, and this in a general-enough case. A system is formally defined as the solution of a boundary-value problem for the set of field equations. Since gravitation propagates with a finite velocity in relativistic theories of gravitation, the relevant boundary-value problem is the initial-value problem. Hence one has to deduce a family of initial data from the initial data valid for the system of interest, S (e.g. the solar system). Then one may expect that the solution fields do admit asymptotic expansions---usually Taylor expansions---as $\lambda\rightarrow 0$. By inserting these expansions into the field equations, one gets each of the latter ones ``split" to successive equations corresponding to the increasing orders in $\lambda$ in the Taylor expansions. The expanded equations, obtained thus, are much simpler than the original equations. The system of interest, S, is assumed to correspond to a small value $\lambda _0 \ll 1$ of $\lambda $, so that one may use the expansions and the expanded equations for S itself---which, of course, is the real aim of the expansion method.

\subsection{Physical framework for the Newtonian limit}

Naturally, the family $(\mathrm{S}^\lambda)$ of systems should be {\it relevant} to the physical situation applying to the system S, which situation justifies the use of asymptotic analysis. In the case of the PNA, the situation is that the gravitational field is weak, and accordingly Newtonian gravity (NG) is supposed to give a very accurate first approximation to the exact behaviour. In less vague terms:
\begin{itemize}
\item {\bf i}) the physical space metric is close to being Euclidean and the local physical time is close to flowing uniformly (equivalently: the physical space-time metric $\Mat{\gamma }$ is close to the flat metric $\Mat{\gamma }^0$); 
\item {\bf ii}) the matter fields are nearly equal to solution fields of the Newtonian equations;
\item {\bf iii}) the Newtonian potential $U_\mathrm{N}$ (or rather some equivalent $V$ of $U_\mathrm{N}$, defined internally to the relativistic theory considered) is {\it small}.\end{itemize}
To make conditions {\bf ii}) and {\bf iii}) precise, it is necessary, as a preliminary step, to define a purely Newtonian situation, in which $U_\mathrm{N}$ itself would be small. In other words, one is led to ask whether a weak-field limit of NG itself could be defined, by considering a family $(\mathrm{S}_\mathrm{N}^\lambda)$ of Newtonian systems. One then finds \cite{FutaSchutz,A23} that there is a {\it similarity transformation} in NG, which is exactly appropriate to that very purpose. If $p^{(1)}$ (pressure), $\mathbf{u}^{(1)}$ (velocity), $\rho^{(1)}=F^{(1)}(p^{(1)})$ (density), and $U_\mathrm{N}^{(1)}$ obey the Euler-Newton equations (the field equations for a perfect fluid in NG), then, for any $\lambda>0$, the fields 
\begin{equation}\label{similarity1}
p^{(\lambda)}(\mathbf{x},T)=\lambda^2p^{(1)}(\mathbf{x},\sqrt{\lambda}\ T), \quad
\rho^{(\lambda)}(\mathbf{x},T)=\lambda\rho^{(1)}(\mathbf{x},\sqrt{\lambda}\ T),
\end{equation}
\begin{equation}\label{similarity2}
U_\mathrm{N}^{(\lambda)}(\mathbf{x},T)=\lambda U_\mathrm{N}^{(1)}(\mathbf{x},\sqrt{\lambda}\ T), \quad
\mathbf{u}^{(\lambda)}(\mathbf{x},T)=\sqrt{\lambda}\ \mathbf{u}^{(1)}(\mathbf{x},\sqrt{\lambda}\ T),
\end {equation}
also obey these equations---provided the state equation for system $\mathrm{S}_\mathrm{N}^\lambda$ is $F^{(\lambda)}(p^{(\lambda)}) = \lambda F^{(1)}(\lambda^{-2} p^{(\lambda)})$. As $\lambda\rightarrow0$, the potential and the density in the bodies decrease like $\lambda$ (while the bodies keep the same size), the velocities decrease like $\sqrt{\lambda}$, and accordingly the time scale increases like $1/\sqrt{\lambda}$. Thus, this similarity transformation defines indeed the weak-field limit in NG itself.

\subsection{Definition of the family of gravitating systems}

Therefore, the idea is to define the initial data of system $\mathrm{S}^\lambda$, in the relativistic theory considered, by applying the similarity transformation (\ref{similarity1})-(\ref{similarity2}) to the initial data for system S. 
\footnote{\ 
Or more precisely, since S is assumed to correspond to a small value $\lambda _0\ll 1$, by applying first the transformation to go from $\lambda_0$ to $\lambda =1$, and second from $\lambda =1$ to the arbitrary value $\lambda$. This amounts to substituting $\zeta  \equiv \lambda /\lambda _0$ for $\lambda $, and $p^{(\lambda _0)}$, etc., for $p^{(1)}$, etc. \cite{A23}.
} 
If conditions {\bf i}) to {\bf iii}) above are asymptotically satisfied as $\lambda \rightarrow 0$, this will undoubtedly define a relevant family $(\mathrm{S}^\lambda)$ of systems. However, one has to find an equivalent of the Newtonian potential in the theory considered. In the scalar ether-theory, condition {\bf i}) means exactly that the scalar field $\psi \equiv -\mathrm{Log}\, \sqrt{\gamma _{00}}$ must tend towards 0 (for {\bf v2}. For {\bf v1} it meant that the field $f\equiv \gamma _{00}$ had to tend towards 1 \cite{A23}). In the asymptotic framework, and in view of the Newtonian similarity transformation, condition {\bf ii}) needs the validity of the following one: 
\begin{itemize}
\item {\bf ii}${\bf '}$) The matter fields in system $\mathrm{S}^\lambda$ have the same order in $\lambda$ as in the transformation (\ref{similarity1})-(\ref{similarity2}).
\end{itemize}
If that condition applies to a family $(\mathrm{S}^\lambda)$, then, by changing the mass and time units for system $\mathrm{S}^\lambda$ in this way: $[\mathrm{M}]_\lambda = \lambda[\mathrm{M}]$ and $\ [\mathrm{T}]_\lambda =
[\mathrm{T}]/\sqrt{\lambda}$, the matter fields $p^{(\lambda)},\ \rho^{(\lambda)}$, and $\mathbf{u}^{(\lambda)}$, become $\mathrm{ord}(\lambda^0)$. Moreover, in these units, we have$\ \lambda \propto 1/c^2$. With the help of these units, one finds easily that, if the field equation (\ref{field}) is to be satisfied by each member of the family, then the field $\psi ^{(\lambda)}$ must be order 1 in $\lambda $, like the Newtonian potential $U_\mathrm{N}$. In this sense, $\psi $ is an equivalent of $U_\mathrm{N}$ in {\bf v2}. Since $\psi $ (unlike $U_\mathrm{N}$) is dimensionless, the most natural field-strength parameter is simply
\be \label{def_lambda}
\lambda \equiv \mathrm{Sup}_{{\bf x} \in \mathrm{M}}\, \psi({\bf x})
\ee
(at the initial time $T=0$, say). The field 
\be
V \equiv c^2 \psi
\ee 
will also be $\mathrm{ord}(\lambda)$ (in fixed units) if $\psi$ is $\mathrm{ord}(\lambda)$, in addition it satisfies the wave equation with the same r.h.s. (in the Newtonian limit where $\sigma \sim \rho$) as Poisson's equation of NG. Since the retardation effects should become negligible in the Newtonian limit, $V^{(\lambda )}$ will be really {\it equivalent} to $U_\mathrm{N}^{(\lambda )}$ as $\lambda \rightarrow 0$. Finally, ${\bf u} = \mathrm{ord}(\sqrt{\lambda })$ (as demanded by condition {\bf ii}${\bf '}$) means a $\lambda^{-1/2}$ dependence of the characteristic time; if one combines this with the fact that $V$ is $\mathrm{ord}(\lambda)$, one expects the asymptotic validity of Eq.~(\ref{similarity2})$_1$ for $V$ :
\be \label{Vequiv}
V^{(\lambda)}({\bf x}, T ) \sim  \lambda V^{(1)}({\bf x}, \sqrt{\lambda}\,T )\ \mathrm{as}\ \lambda \rightarrow 0.		                   
\ee
In uniform conditions, one may differentiate (\ref{Vequiv}) with respect to $T$: 
\be
\partial _T V^{(\lambda)}({\bf x}, T ) \sim  \lambda^{3/2} \partial _T V^{(1)}({\bf x}, \sqrt{\lambda}\,T )\ \mathrm{as}\ \lambda \rightarrow 0.		                   
\ee
Therefore, in order to satisfy conditions {\bf i}), {\bf ii}${\bf '}$) and {\bf iii}), it is most natural to define the family $(\mathrm{S}^\lambda)$ by the following family of initial conditions:
\begin{equation}\label{PN_IC1}
p^{(\lambda)}(\mathbf{x})=\lambda^2p^{(1)}(\mathbf{x}), \quad
\rho^{\ast (\lambda)}(\mathbf{x})=\lambda\rho^{\ast (1)}(\mathbf{x}),
\end{equation}
\begin{equation}\label{PN_IC3}
\mathbf{u}^{(\lambda)}(\mathbf{x})= \sqrt {\lambda }\, \mathbf{u}^{(1)}(\mathbf{x}),
\end {equation}
\begin{equation}\label{PN_IC2}
V^{(\lambda)}(\mathbf{x})=\lambda V^{(1)}(\mathbf{x}), \quad\partial_TV^{(\lambda)}(\mathbf{x})=\lambda^{3/2} \partial _TV^{(1)}(\mathbf{x}).
\end {equation}
($\rho ^\ast$ is the proper rest-mass density.) We may confidently expect that, in a significant interval of the ``dynamical time" \cite{FutaSchutz} $\tau \equiv T\sqrt{\lambda }$,  the solution fields will remain of the same order in $\lambda $ as in this initial data, hence conditions {\bf i}), {\bf ii}${\bf '}$) and {\bf iii}) will indeed be satisfied in the asymptotic sense, in that interval. 

\subsection{Implementation of the method in the scalar theory and in GR} \label{Implementation}

Using the varying units $[\mathrm{M}]_\lambda = \lambda[\mathrm{M}]$ and $\ [\mathrm{T}]_\lambda = [\mathrm{T}]/\sqrt{\lambda}$, the matter fields {\it and} the scalar gravitational field $V$ become $\mathrm{ord}(\lambda^0)$, while the small parameter $\ \lambda \propto 1/c^2$. This makes the expansions straightforward. {\it All fields} (of course) depend on $\lambda$ and are Taylor-expanded \{at least up to some order $n_\mathrm{max}$, beyond which other expansion functions might have to be used, as done in GR \cite{ChandraEsposito, Anderson-et-al82}\}: 
\be  
\rho = \sum_{k=0}^n \rho_k  \lambda^k+ O(\lambda^{n+1}), \quad \mathrm{etc.\quad(and}\ \lambda = 1/c^2)
\ee\\
Inserting these expansions into the exact equations, each of the latter ones splits to $n+1$ exact equations (it is just coefficient identification and remains true if the expansion basis is larger). Hence, the numbers of the independent equations and the independent unknowns remain equal. The zero-order equations (corresponding to $k=0$) turn out to be the Euler-Newton equations of Newtonian gravity. Thus, the Newtonian limit is satisfied, in particular it is so for condition {\bf ii}). The $(k=1)$--equations give the first PN (1PN) correction, which is linear in the 1PN fields, like $\rho _1$. The details can be found in Ref. \cite{A23}, where the method is applied to {\bf v1}. The modification for {\bf v2} is straightforward.\\
 
In GR, a method similar to that one which we have just presented, has been introduced in a particular case by Futamase \& Schutz \cite{FutaSchutz} (although the present author came to this method independently \cite{A23}). Namely, their initial condition for the matter fields was the same as here, Eqs.~(\ref{PN_IC1})-(\ref{PN_IC3}), but their initial condition for the space metric and its time-derivative, playing the role of Eq.~(\ref{PN_IC2}) here, was their Eqs.~(3.13)$_{4-5}$, which are in our notation: 
\be \label{ICmetric-F&S}
\sqrt{-\gamma }\ \gamma ^{ij} ({\bf x}) = \delta _{ij}, \qquad
\left(\sqrt{-\gamma }\ \gamma ^{ij} \right)_{,0} ({\bf x}) =0 \qquad (1\leq i\leq 3,\quad 1\leq j\leq 3).
\ee
This condition is verified when the spatial metric has the ``conformally-Euclidean" (or isotropic) form
\be \label{spatial-F&S}
g_{ij} = \sqrt{-\gamma }\ \delta _{ij}, 
\ee
in fact it is just the condition that one would impose if one would wish to have the space metric (\ref{spatial-F&S}), without imposing any a priori restriction on the factor $\sqrt{-\gamma }$. This initial condition does not directly impose that the gravitational field is weak, since (\ref{ICmetric-F&S}), in contrast with (\ref{similarity2})$_1$ and (\ref{PN_IC2}), does not depend on the small parameter $\lambda $. This seems a priori surprising. Thus, in the harmonic gauge, assuming the metric (\ref{spatial-F&S}) together with the Newtonian-limit conditions for the matter fields, (\ref{PN_IC1}) and (\ref{PN_IC3}), constrains the solution of the Einstein equations to be a weak field. In the scalar theory also, we used the field equation to get condition (\ref{PN_IC2}), but things look less straightforward in GR. In the framework of GR, it is legitimate to assume the initial condition (\ref{ICmetric-F&S}), but it looks quite particular: there are many, very different, possibilities, and one would like to know what would happen if he would choose very different initial conditions for the space metric, e.g. ones corresponding to an ``anisotropic" space metric. (We shall come back to this question in Subsect. \ref{caseGR}.) 
\footnote{\ 
When trying to implement an asymptotic PNA scheme for GR, there appear two basic difficulties as compared with the present theory, which is much simpler than GR. One arises from the fact that the gravitational field is then a space-time tensor, thus has much more freedom than in Newton's theory, so that a priori the Newtonian limit might be defined in different ways. The other difficulty is due to the fact that the $(\mu 0)$ components of the Einstein equations $G^{\mu \nu }=\kappa T^{\mu \nu }$ are not evolution equations, and instead impose four nonlinear constraints on the initial data for the metric \cite{Weinberg}. Depending on the gauge condition adopted, it may be more or less difficult to account for these ``constraint equations" in the definition of the initial data. In Ref. \cite{FutaSchutz}, the harmonic gauge was used. In that well-known gauge, the Einstein equations, thus including the constraint equations, take a relatively simple form, but still they remain nonlinear. As a matter of fact, the initial conditions for the $\sqrt{-\gamma }\ \gamma ^{\mu 0}$ components (which conditions have to hold as a consequence of (\ref{ICmetric-F&S}) and the Einstein equations) were not solved explicitly in Ref. \cite{FutaSchutz}. 
}
Finally, Futamase \& Schutz \cite{FutaSchutz} derived the local equations in a form which is not explicit enough for further use, and derived no ``global" equations i.e. no equations of motion for the mass centers of extended bodies. 

\subsection{Comparison with the standard PNA scheme used in GR} 

An extremely good agreement is found between astronomical observations and calculations based on GR \cite{Will}. However, such calculations use the standard PNA scheme, which remains essentially identical to that one which was proposed by Fock \cite{Fock59} and Chandrasekhar \cite{Chandra65}. In the standard scheme, no family of systems is envisaged; $1/c^2$ is {\it formally} considered as a small parameter; and the matter fields $p, \rho , {\bf u}\ $ are {\it not} expanded. Since, in an asymptotic framework, the matter fields do depend on the small parameter, it follows that the splitting of each exact equation into $n+1$ equations would not be justified in the standard scheme. The problem is that this splitting is needed at least for the equations for the gravitational field \cite{O1,A23}. Detailed arguments, and references to more recent literature, can be found there \cite{O1,A23}. 

\section{Equations of motion of the mass centers}
The mass centers of the celestial bodies are defined \cite{A25} as local barycenters of the {\it rest-mass density}. Therefore, to get the PN equations of motion of the mass centers (EMMC's), the local PN equations of motion [which are just the expansion of Eq.~(\ref{continuum}), specialized to a perfect fluid] are integrated in the domain $\mathrm{D}_a$ occupied by body $(a)$ \cite{A25}. One finds that {\it the internal structure of the bodies (e.g. the density and velocity profiles inside them) influences the motion} already from the first PNA \cite{A32}. This follows naturally \cite{A32} from using the ``asymptotic" method of PNA and, in the author's opinion,  should hold true for GR if an asymptotic PNA was used there.

\section{Point-particle limit of the EMMC's and the WEP} \label{PointParticleLimit}
Once it had been recognized \cite{A32} that the internal structure of the gravitating bodies does influence their motion, it was important to check whether or not this influence cancels in the limit of a very small body. If it does not, the weak equivalence principle (WEP) is violated.

\subsection{Framework for the point-particle limit}

We consider a system of $N$ bodies $(1), ..., (N)$, that move under the 1PN gravitational field produced by them all. (Since, as explained in Subsect. \ref{Implementation} above, the asymptotic 1PN approximation leads to split each of the exact local equations to two equations, the local equations of the first PNA make a closed exact system \cite{A25}. Hence, assuming that the physical system under study is described accurately enough by the first PNA, we do not need to consider the weak-field parameter $\lambda$.) The size of one of the bodies, say $(1)$, is a small parameter $\xi$, with $\xi\rightarrow 0$, thus a family $(\mathrm{S}^\mathrm{\xi})$ of 1PN systems is considered \cite{A33}. The initial data is independent of $\xi $ apart from the size of the small body. One just has to find the main terms in the expansions as $\xi\rightarrow 0$ of the integrals \cite{A25} that enter the general form of the EMMC's. This is a bit long and technical but not really difficult \cite{A33}.

\subsection{Result of the point-particle limit}

The point-particle limit of the (1PN) EMMC's {\it of {\bf v1}} was taken \cite{A33}. It was found that a structure-dependent part of the acceleration, ${\bf A}_S$, remains at the point-particle limit $\xi \rightarrow 0$. Moreover, in the static spherically symmetric (SSS) case, the limit value of the acceleration as $\xi\rightarrow 0$ differs from the acceleration of a test particle just by ${\bf A}_S$:
\be \label{A-limit-SSS}
\lim_{\xi \rightarrow 0} {\bf A}^\xi = {\bf A}_S + (\mathrm{1PN \ acceleration\ of\ a\ test\ particle\ in\ the\ SSS\ field}).
\ee
Thus, {\it the Weak Equivalence Principle is violated {\it in {\bf v1}}.} This was a bad surprise, because the validity of the WEP for {\it test particles} and even for a ``test continuum" was built in the very construction of the theory, see Eqs.~(\ref{Newtonlaw}) and (\ref{continuum}) above. Moreover, this violation seemed very likely to {\it  kill} {\bf v1}, since, for instance, $\abs{{\bf A}_S}$ is of the order of $10^5$ times the anomalous residual acceleration found \cite{AndersonJPL98,AndersonJPL02} for the Pioneer 10 spacecraft...\\

Therefore, the reason for the violation was carefully analysed \cite{A33} (see Subsect. \ref{ReasonsViolWEP} below for a summary), and {\bf v2} was constructed in order to avoid this violation. And indeed the violation is found to not take place (at the 1PN approximation) in {\bf v2}: we have ${\bf A}_S = {\bf 0}$ in the {\it general case} for {\bf v2}, {\it this has been checked}. Moreover, Eq.~(\ref{A-limit-SSS}) still holds (thus with ${\bf A}_S = {\bf 0}$) for {\bf v2}. I.e., for {\bf v2}, the acceleration of an extended body (1) in the SSS field produced by a massive body (2) coincides, as closely as desired, with the acceleration of a test particle in the same SSS field, provided that body (1) is small enough.

\subsection{Reasons for the WEP violation in v1 and for no violation in v2} \label{ReasonsViolWEP}

The general reason which makes it {\it a priori possible} that the WEP might be violated for extended bodies is simply {\it the non-linearity of the theory!} Indeed, any kind of material object must contribute to the gravitational field. In Newton's theory, the gravitational attractions between the particles of a body sum to zero (and the same is true for the corresponding torques), due to actio-reactio equality. (It can also be checked on the continuous form of Newton's attraction law, based on the Poisson equation, that the resultant of the ``self field" created by a body is zero.) However, the actio-reactio equality cannot even be {\it formulated} in a nonlinear theory, because one cannot define what is the ``gravitational attraction exerted by an object over another object". Then, the body contributes nonlinearly to the gravitational field and it cannot be guaranteed in advance that the body's contribution to its own acceleration will decay with the body's size.
\\

Moreover, the influence of the structure of the bodies on the gravitational field, hence on the motion, is to be generally expected in any relativistic theory, due to the mass-energy equivalence. The asymptotic PNA separates the equations of motion of the different orders. This makes the structure influence occur explicitly in the general equations for the 1PN corrections to the motion of the mass centers, because these equations involve several integrals of the Newtonian matter fields \cite{A25}. Again, it is not a priori obvious that this structure influence must cancel in the point-particle limit.
\\

But the specific reason that makes the WEP violation {\it actually occur} in {\bf v1} is the presence in {\bf v1}'s PN spatial metric $\Mat{g}_{(1)}$ of the spatial derivatives $U_{,i}$ of the Newtonian potential. Indeed, the local equation of motion depends of course on the Christoffel symbols of the metric, thus, in that case, it depends on the second derivatives $U_{,i,j}$. The equation of motion of the mass center ${\bf a}$ of body $(1)$, being got by integration of the local equation inside the volume of $(1)$, depends on an integral involving the $U_{,i,j}$'s. Now the self part of $U_{,i,j}$ [i.e., the contribution of (1) to $U_{,i,j}$] does not depend on the size $\xi$ of (1), hence does not evanesce with $\xi$. This is the reason why the acceleration ${\bf \ddot{a}}$ depends on the structure of the small body (1) itself, even in the limit $\xi \rightarrow 0$ \cite{A33}.\\

Since, in {\bf v2}, the spatial metric is isotropic [Eq.~(\ref{spacemetric})], its PN expansion involves the Newtonian potential $U$ (precisely: $U\equiv V_0$, which obeys indeed the Poisson equation), but it does not involve its spatial derivatives $U_{,i}$. Therefore, the WEP should not be violated any more in the point-particle limit of the 1PN EMMC's in {\bf v2}. However, the scalar field equation and the metric being different from {\bf v1}, this had to be checked carefully. This has been done, and indeed the WEP applies (at the 1PN approximation) to {\bf v2}.

\subsection{The case with general relativity} \label{caseGR}
Now the $U_{, i}$'s, which are responsible for the WEP violation in {\bf v1} of the scalar ether-theory, are also there in the PN approximation to the standard form of Schwarzschild's metric. The latter is also spatially anisotropic, in just the same way as is the metric of {\bf v1} (i.e., the direction ${\bf g}$, here the radial direction, is preferred). In fact, the standard Schwarzschild metric is the unique solution of {\bf v1} for the SSS case (under the usual condition that the behaviour is Newtonian at infinity)---see e.g. Ref. \cite{A34}, Subsect. 2.1. \\

Hence, it is natural to ask the question: what about GR? In the usually-adopted harmonic gauge, the {\it standard} 1PN approximation of the spatial metric is isotropic and contains $U$ but not the $U_{,i}$'s, hence no WEP violation does occur. Now in the frame of an {\it asymptotic} PNA scheme, imposing an initial condition corresponding to an isotropic metric is consistent with the harmonic gauge \cite{FutaSchutz}, but it should be tried to impose different conditions (Subsect.~\ref{Implementation}). Moreover, gauges do exist, in which the standard Schwarzschild solution is the unique solution of the SSS case with Newtonian behaviour at infinity (e.g. one is shown in Ref. \cite{O2}, Note 15; a more convenient one is also known to the author). In a such ``Schwarzschild-like gauge", the spatial metric of the {\it general case} would be expected to be ``anisotropic", and its PN approximation would be expected to involve derivatives $U_{, i}$. Hence, it seems quite likely that the WEP could also be violated in GR with a such gauge. In GR, then, the validity of the WEP for a small extended body at the 1PN approximation might depend on the gauge condition.

\section{Conclusion}

A scalar theory with a preferred reference frame has been summarized. A new version (``{\bf v2}") has been presented. In {\bf v1}, the space metric was anisotropic, whereas in {\bf v2} it is isotropic. The complete metric of {\bf v2}, Eq.~(\ref{spacetimemetric}) with (\ref{spacemetric}), is an often-met one \cite{Ni72,Podlaha00,Broekaert03}. The  formal specificity of the present theory (independently of its motivation and its heuristic) is in its revendicated preferred-frame character (Subsect. \ref{PRF}), its dynamics (Subsect. \ref{Dynamics}), and its scalar field equation implying a local energy conservation (Subsect. \ref{field&energy}). 
\\

To test that theory in celestial mechanics, an ``asymptotic" PN 	scheme, as in applied mathematics, was developed. The resulting equations of motion for a self-gravitating system of extended	bodies include {\it internal-structure effects}. For {\bf v1} {\it only}, the internal-structure influence subsists at the point-particle limit. This is a {\it severe violation of the WEP}, although the WEP is satisfied for test particles. {\bf v2} was constructed precisely to avoid this violation---and indeed it {\it does avoid it}.
\\ 

Using an ``asymptotic" approximation scheme, the same WEP violation as in {\bf v1}  might occur in GR also, e.g. in a gauge where the SSS solution is the standard Schwarzschild metric, and/or with an initial condition corresponding to an anisotropic space metric---but it would be more difficult to check it than in the present theory, due to the greater complexity of GR.
\\

{\bf Acknowledgement.} I am grateful to the organizers of GAS '04, in particular to P. Fiziev, R. Rashkov, and M. Todorov, for their warm hospitality and for the good organization of this Workshop.

\end{document}